\documentclass[a4paper,11pt]{article}
\pdfoutput=1 % if your are submitting a pdflatex (i.e. if you have
\usepackage{jinstpub} % for details on the use of the package, please
 \usepackage{siunitx}
\usepackage{xargs}   
\usepackage{soul}
\usepackage{siunitx}
\newcommandx{\unsure}[2][1=]{\todo[linecolor=red,backgroundcolor=red!25,bordercolor=red,#1]{#2}}
\newcommandx{\change}[2][1=]{\todo[linecolor=blue,backgroundcolor=blue!25,bordercolor=blue,#1]{#2}}
\newcommandx{\info}[2][1=]{\todo[linecolor=OliveGreen,backgroundcolor=OliveGreen!25,bordercolor=OliveGreen,#1]{#2}}
\newcommandx{\improvement}[2][1=]{\todo[linecolor=Plum,backgroundcolor=Plum!25,bordercolor=Plum,#1]{#2}}
\newcommandx{\thiswillnotshow}[2][1=]{\todo[disable,#1]{#2}}

\title{\boldmath Temperature-dependent fluorescence emission spectra of acrylic (PMMA) and tetraphenyl butadiene (TPB) excited with UV light}

\author[a]{J.~M.~Corning,}
\author[b,1]{G.~R.~Araujo\note{Current address: Physik Institut der Universit{\"a}t Zürich, CH-8057 Zurich, Switzerland},}
\author[a,2]{P.~C.~F.~Di~Stefano\note{Corresponding author.},}
\author[a]{V.~Pereymak,}
\author[b]{T.~Pollmann,}
\author[a]{and P.~Skensved}

\affiliation[a]{Department of Physics, Engineering Physics \& Astronomy, Queen's University,\\Kingston, Ontario, Canada, K7L 3N6}
\affiliation[b]{Physik-Department E15, Technische Universit\"{a}t M\"{u}nchen, Garching, Germany}

\emailAdd{distefan@queensu.ca}

\abstract{Acrylic (poly(methyl methacrylate) or PMMA) is commonly used as a vessel to hold scintillating liquids in rare-event searches.  Certain types of PMMA can fluoresce with a low efficiency at room temperature. %\todo{I would soften this sentence. The fluorescence of PMMA is not widely known and PMMA is not a fluorescent material in general. Furthermore, it is not clear if the fluorescence is due to impurities. It could also be due to color centers and defects.}.  
We have investigated the fluorescence spectra under \SI{280}{\nm} ultraviolet (UV) excitation of the PMMA used in the DEAP dark matter search at various low temperatures.   Fluorescence of this PMMA is observed, with an increasing intensity as the sample is cooled from a temperature of 300~K down to 4~K. The common wavelength-shifter tetraphenyl butadiene (TPB) is also measured for use as a reference.}

\proceeding{LIDINE: Light Detection In Noble Elements\\
  2019\\
  Manchester UK}

\begin{document}
\maketitle
\flushbottom

%\todo[color=pds,inline]{5 pages allocated (not counting, title, abstract, references) --- deadline Nov 22}
%\todo{5 pages, so we do not have much space left}
\section{Introduction}
\label{sec:intro}

Several rare-event search experiments use a scintillating target contained in support structures made from plastics. The DEAP-3600 experiment, for instance, uses an acrylic vessel to contain the liquid argon (LAr) target~\cite{DEAP}. Acrylic, or poly(methyl methacrylate) (PMMA), has been reported to fluoresce at a low intensity when excited by ultraviolet (UV) light (\cite{fluores_paper} and references therein). The unexpected fluorescence of construction and support materials, even at a very low efficiency, %\todo{changed level-> intensity, since we also did that for the last paper}, 
could be a concern for those experiments that look for rare events at low energies. For this reason, we have previously investigated the fluorescence response of acrylic and polytetrafluoroethylene (PTFE) to vacuum UV photons from \SIrange{130}{250}{\nano\meter} at room temperature~\cite{fluores_paper}. These wavelengths span the range of LAr and LXe scintillation light, as well as the far UV and part of the mid-UV spectral range of Cherenkov light created within the acrylic. However, the samples were not measured at the corresponding cryogenic temperatures of \SI{87}{\K} (LAr) or \SI{165}{\K} (LXe), even though the luminescence of a material can depend on its temperature~\cite{scint_TPB}. No fluorescence was observed, so~\cite{fluores_paper} gives limits on the level of fluorescence of the measured samples at room temperature.
%
%The fluorescence response of a material can depend on its temperature.

In view of a future comparison of the fluorescence of PMMA to a well-known reference, we have also measured the low-temperature spectra of
%The response of the plastic is compared \todo{is it? there is no sentence or number in the analysis comparing both} to the response of 
the wavelength shifter 1,1,4,4-tetraphenyl-1,3-butadiene (TPB). TPB is commonly used to shift liquid noble gas scintillation light to the blue spectral range, so that it is easier to detect. In multiple experiments, such as DEAP-3600, is it used in combination with acrylic~\cite{DEAP}. Therefore, it is a good reference for wavelength shifting efficiencies. In a first step toward characterizing the potential fluorescence of acrylic and PTFE at cryogenic temperatures, we have measured the response of acrylic to \SI{280}{\nano\meter} excitation photons at temperatures from \SI{300}{\K} down to \SI{4}{\K}.

\section{Samples, Procedures and Set-up}

The principle of these measurements was to excite the samples with continuous \SI{280}{\nm} light and measure their fluorescence emission spectra at temperatures from \SIrange{300}{4}{\K}.
Since the UV photons do not penetrate far into the material, the treatment of the surface can have a large effect on the fluorescence response. Furthermore, the response can be modified by impurities, additives, and by the conditions under which the material was machined or otherwise treated. Therefore, we use a sample of acrylic from the same batch the DEAP-3600 acrylic vessel was made from. The sample is the same as that described in \cite{fluores_paper} as `RPT acrylic,' supplied to the DEAP-3600 Collaboration by Reynold's Polymer Technology.
The TPB sample was vacuum evaporated onto a piece of generic acrylic of the same size as the RPT acrylic sample. The thickness of the deposited TPB layer was measured to be \SI[separate-uncertainty = true]{0.35(3)}{\micro\meter}.%change to 

%Since the fluorescence response of PMMA can be attributed to microscopic impurities, additives, or manufacturing flaws\color{blue}$^{1}$\color{black}, the sample studied in this work was made from the same batch of acrylic used to construct the DEAP-3600 acrylic vessel, made by Reynold's Polymer Technology (henceforth RPT acrylic). The TPB sample was vacuum evaporated on acrylic.~\footnote{For technical details about machining and cleaning of the samples, see~\cite{fluores_paper}.}
%The RPT acrylic sample was cut into a round disc with \SI{25}{\mm} diameter, ??\SI{}{\mm} \todo[backgroundcolor=blue!20!white, linecolor=blue!20!white]{Measure thickness of TUM RPT.}thick. Another sample of generic acrylic identical in size was made, and TPB was vacuum evaporated onto one side. The thickness of the deposited TPB layer was measured to be ??\SI{}{\micro\meter}.

\begin{figure}[htp]
    \centering
    \raisebox{0.17\height}{\includegraphics[width=0.25\textwidth]{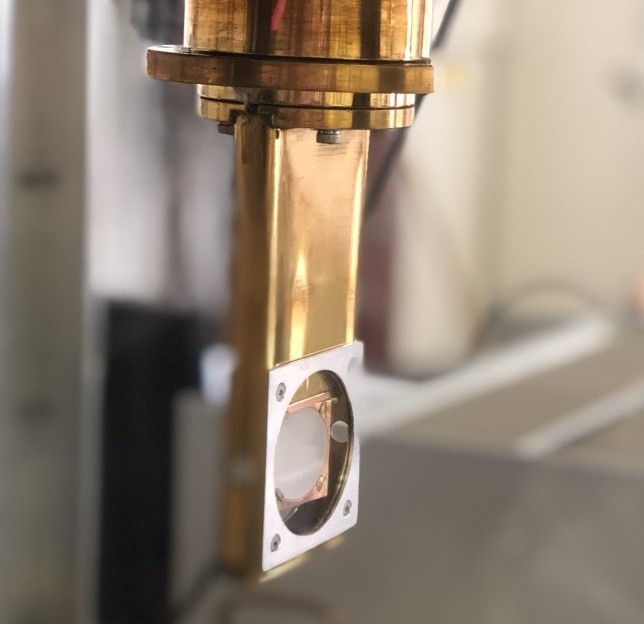}}
     \includegraphics[width=0.65\textwidth]{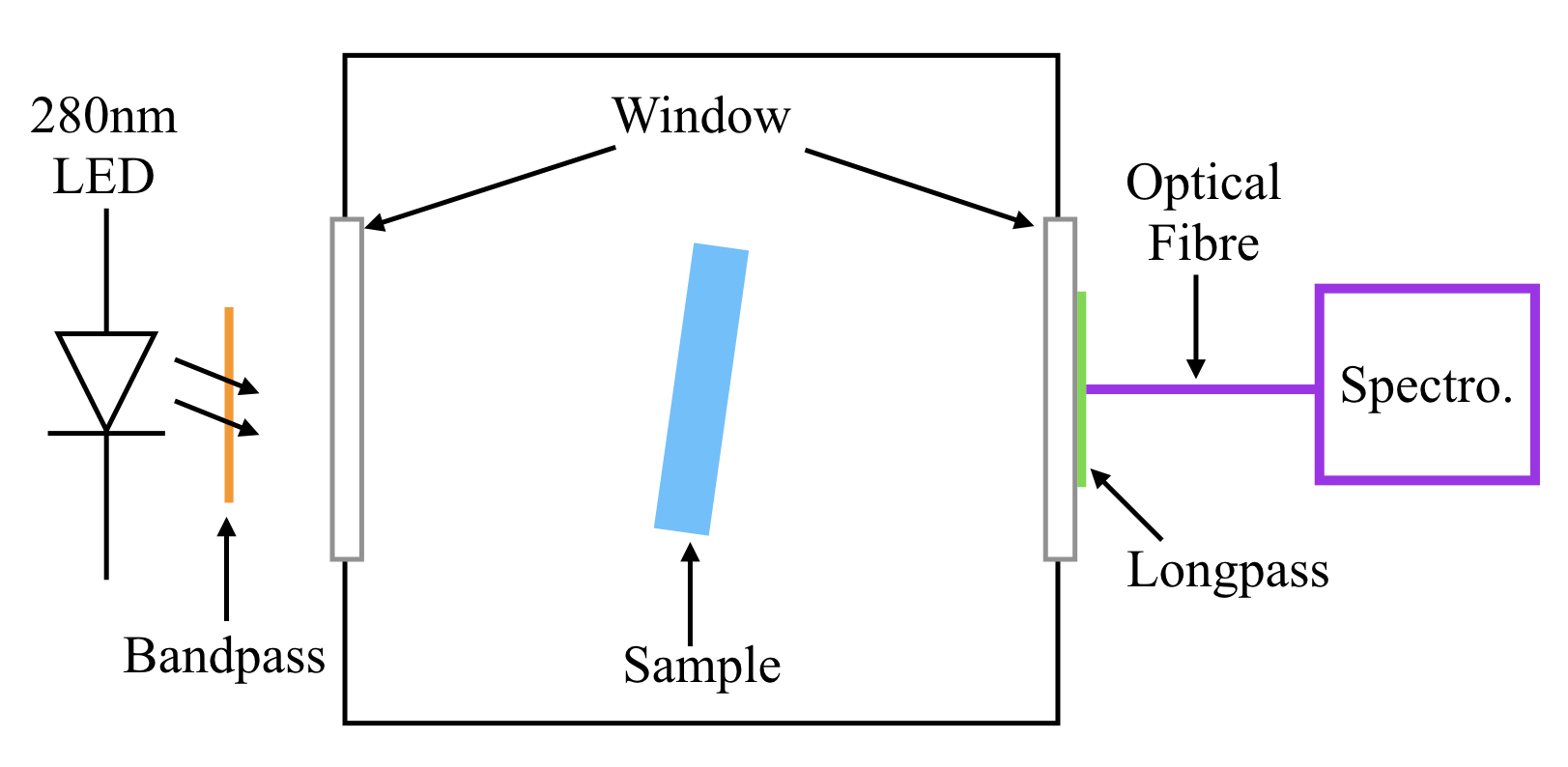}
   \caption{\textit{Left: }TPB coated sample installed into the copper sample holder, which is then attached to the cold finger of the cryostat. All of this is shown enclosed in the inner-most cryostat chamber. \textit{Right: }Top-down cross-sectional view of the schematic used for this experiment. The light shines from the LED through the bandpass filter, three cryostat windows (only one window is illustrated) %(physically, there are 3 windows - one for each cryostat chamber as explained in the text), 
   and reaches the sample. If the sample fluoresces at visible wavelengths, the fluorescence light goes through the adjacent windows, longpass filter, and then into the optical fibre. In the case of the TPB sample, the TPB coating is on the side facing the LED.}
    \label{fig:setup}
\end{figure}

The samples were installed into a custom copper sample holder which rotated the sample approximately \SI{8}{\degree}, relative to the incoming light beam, around the vertical axis. %\footnote{This rotation was chosen to also permit excitation of sample from the side which is observed. \textcolor{purple}{Unclear what this means.}}.  
The holder was then attached onto the cold finger of the cryostat, as shown in Figure~\ref{fig:setup}.
The cryostat is closed-cycle, with a compact optical geometry to optimize light-collection efficiency~\cite{cryo_details}. The cold finger is enclosed within a vacuum shroud and two radiation shields.  The sample can be viewed from the outermost shroud through a corresponding series of Spectrosil 2000
fused silica windows, which have more than 90\% transmission at the wavelengths of interest~\cite{cryo_details}.%\todo{only necessary in case we discuss why we measured only with 280nm light and not lower}\st{These windows have a low transmittance at VUV light wavelengths, such as the characteristic LAr scintillation wavelength}

A schematic of the experiment is shown in Figure \ref{fig:setup}. A commercial \SI{280}{\nm} LED was attached to the side of the outer cryostat chamber.
%, roughly perpendicular to the surface of the sample. 
When the TPB-coated sample was installed, the LED was shining onto the coated surface; when the RPT acrylic sample was installed, the LED was shining onto the smoother finished surface. The LED was powered in direct-current (DC) mode, and used with an Edmund Optics Hard Coated \SI{280}{\nm} bandpass filter to ensure that only monochromatic \SI[separate-uncertainty = true]{280(5)}{\nm} light reached the samples. We used the Horiba Scientific Symphony II CCD Detector and MicroHR Automated Imaging Spectrometer for spectral imaging. The optical fibre of the spectrometer was attached to the opposite window of the outer chamber, facing the opposite surface than the LED was shining onto. A longpass filter made of transparent, UV-absorbing (UVA) acrylic was installed in front of the optical fibre, to eliminate any UV light that may have reached the fibre's position through or around the sample.

The optical cryostat was put under vacuum for all measurements. An external CryoCon 24 temperature controller system was used to set and monitor the cold finger's temperature. All external components were at room temperature. Spectra were taken in descending order through temperatures \SI[separate-uncertainty = true]{300.0(10)}{\kelvin}, \SI[separate-uncertainty = true]{163.0(15)}{\kelvin} (LXe), \SI[separate-uncertainty = true]{87.0(15)}{\kelvin} (LAr), and \SI[separate-uncertainty = true]{4.0(1)}{\kelvin}. 

%For internal references use label-refs: see section~\ref{sec:intro}. Bibliographic citations can be done with cite: refs.~\cite{a,b,c}.

%A similar solution is available for figures via the \texttt{subfigure} package (not loaded by default and not shown here).All figures and tables should be referenced in the text and should be placed on the page where they are first cited or in subsequent pages. Positioning them in the source file after the paragraph where you first reference them usually yield good results. See figure~\ref{fig:i} and table~\ref{tab:i}.

%\begin{figure}[htbp]
%\centering % \begin{center}/\end{center} takes some additional vertical space
%\includegraphics[width=.4\textwidth,trim=30 110 0 0,clip]{example-image-a} 
%\qquad
%\includegraphics[width=.4\textwidth,origin=c,angle=180]{example-image-b}
% "\includegraphics" from the "graphicx" permits to crop (trim+clip)
% and rotate (angle) and image (and much more)
%\caption{\label{fig:i} Always give a caption.}
%\end{figure}

\section{Analysis and results}

Spectra from the sample were corrected for the spectrum from a dark exposure of equal exposure time. 
%For each spectrum of light collected, the spectrometer recorded a dark window of equal exposure time, and subtracted this dark exposure before outputting the final collected spectrum. This allows us to be confident that internal spectrometer components are not having an effect on our data. 
The scanning range of one window of the spectrometer is approximately \SI{285}{\nm}, so multiple windows were recorded at each temperature to scan from the approximately \SI{350}{\nano\meter} to the near infrared. % We have the longpass filter so we're not reading re-emission down to the LED wavelength

% (TP) I think we decided to drop this. To apply a response correction to the spectrometer data, we measured the RT emission spectrum of a replica TPB sample\todo[color=pds]{Was this the same sample?}\todo{It was a replica of the sample. Evaporated at the same time, for the same amount of time and at a similar position. Also on the same substrate.} with a Cary Eclipse Fluorescence Spectrophotometer, already corrected for the response of that spectrometer, as the true TPB spectrum, from \SIrange{375}{600}{\nm}. The ratio of the normalized true spectrum to our normalized room temperature spectrum (at \SI{300}{K}) gives the correction function of our spectrometer in this range. This correction function was then applied to the spectra of TPB and RPT at all measured temperatures. All spectra were re-binned into wider wavelength bins in order to better show the overall shape of the spectra.

\begin{figure}[htp]
    \centering
    \includegraphics[scale=0.4]{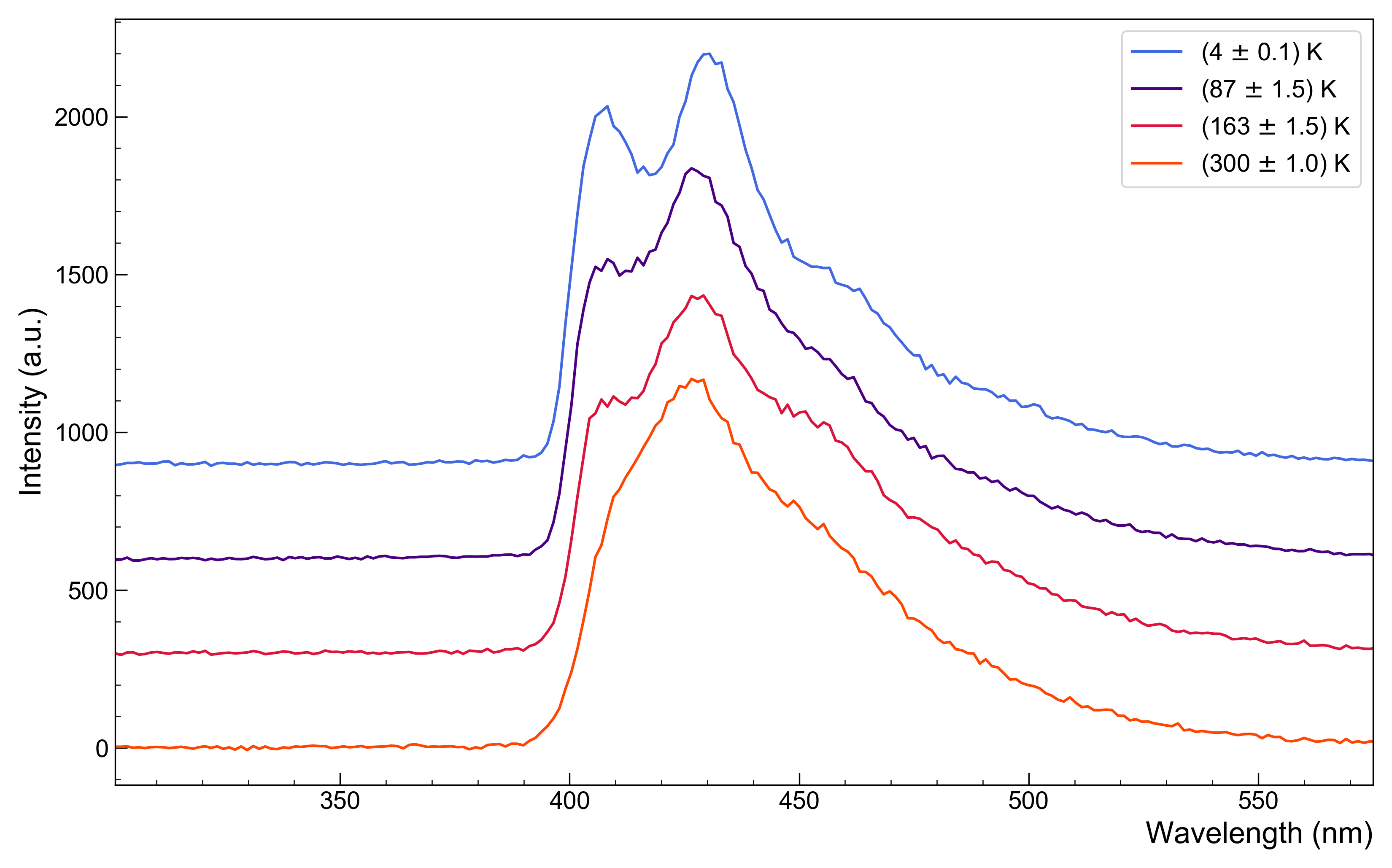}
    \caption{Spectra of the TPB sample over all measured temperatures. An arbitrary vertical offset was applied to better show differences in the spectral shapes.}
    \label{fig:tpb_spectra}
\end{figure}

%\begin{equation}
%    C(\lambda) = \frac{I_{true}(\lambda)/I_{true, max}}{I_{meas.}(\lambda)/I_{meas., max}}
%\end{equation}
%\todo[color=pds]{What does this equation represent, and where is it referred to?}
Figure~\ref{fig:tpb_spectra} shows the spectra of TPB over all measured temperatures. The intensity shows an mildly increasing trend from \SIrange{300}{4}{\K} over the main emission peak from \SIrange{375}{450}{\nm}. An additional peak not seen at room temperature, at approximately~\SI{400}{\nm}, also increases as temperature decreases. This feature has been seen in previous measurements of TPB films down to LAr temperature, and has been attributed to vibrational structures in the molecules~\cite{tpbfilms}.

\begin{figure}[htp]
    \centering
    \includegraphics[scale=0.4]{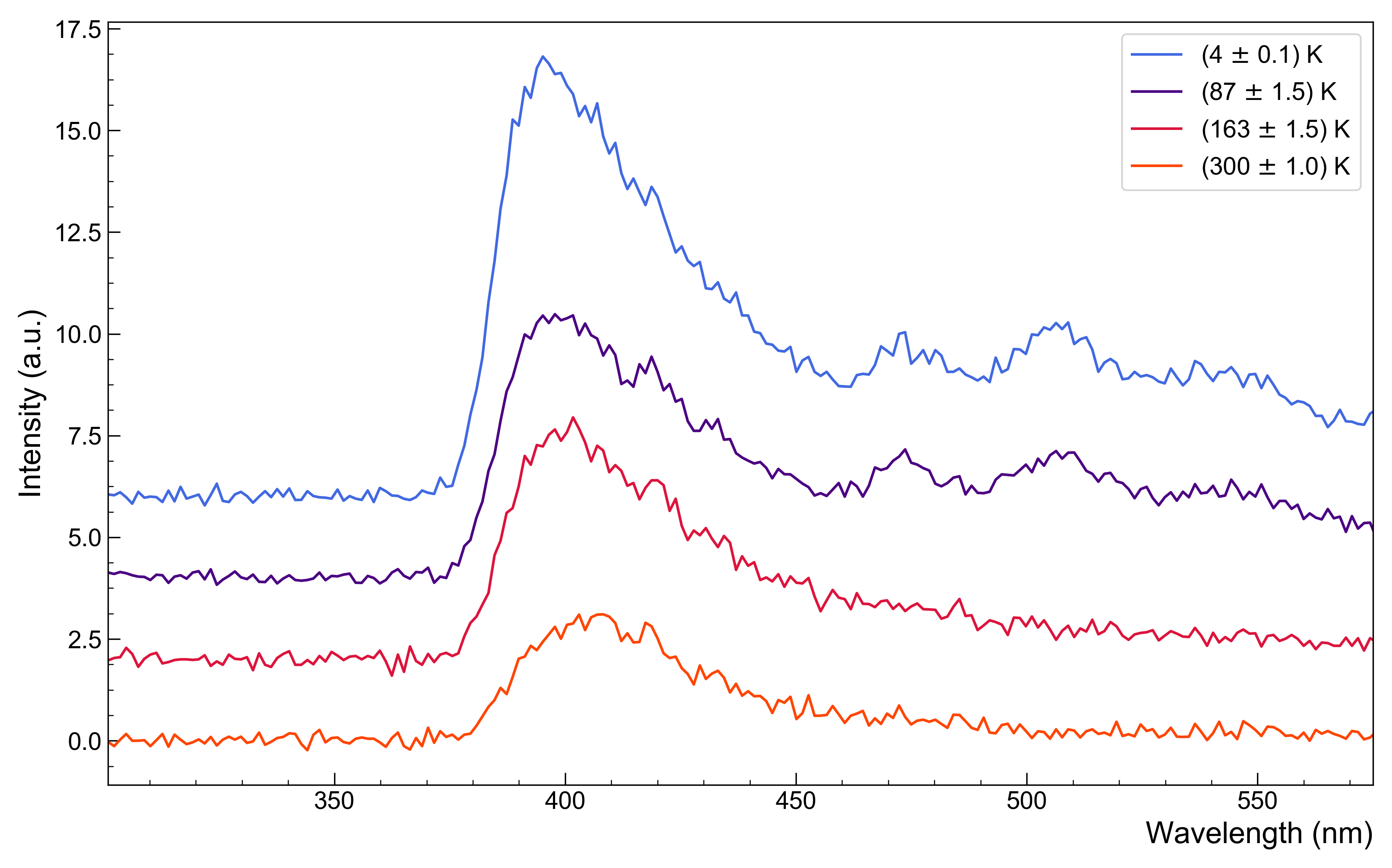}
    \caption{Spectra of the RPT acrylic sample over all measured temperatures. An arbitrary vertical offset was applied to better show differences in the spectral shapes.}
    \label{fig:rpt_spectra}
\end{figure}

The emission spectra of RPT acrylic are shown in Figure~\ref{fig:rpt_spectra}. We detect visible fluorescence light from the sample well above the noise level of the spectrometer between approximately \SI{360}{\nano\meter} and \SI{450}{\nano\meter}, and at a lower level over \SIrange{450}{600}{\nano\meter}. On the shorter-wavelength side, the structure may be cut off by the longpass filter. The intensity of the emission feature increases as the temperature decreases, particularly in the main emission peak around \SI{400}{\nm}.  A dedicated room-temperature measurement was carried out with and without the acrylic sample which confirmed that the fluorescence originates from the sample, as opposed to other components of the setup.
%\todo[color=pds,inline]{If the correction is the one in appendices, it is quite noisy.}

\section{Conclusion}

We measured the temperature-dependent emission spectra of TPB and RPT acrylic excited with \SI{280}{\nano\meter} light. The TPB spectra are consistent with previous results at higher temperatures~\cite{tpbfilms}, showing more clearly the vibrational structures at decreasing temperatures. The fluorescence emission spectrum of this batch of acrylic at low temperatures was measured for the first time, and its intensity increased as temperature decreased. This effect could be due to the decrease of non-radiative transitions at low temperatures, or decrease of quenching of fluorescence \cite{fluorbook}. 

On the usage of acrylic in a detector for rare-events, this fluorescence can shift UV Cherenkov light produced in acrylic to the visible range. Since the UV part of the Cherenkov spectrum is not otherwise transmitted through the acrylic to be detected by the light sensors, this wavelength-shifting response must be included in detector Monte-Carlo simulations to accurately model the detector and expected backgrounds. 

Fluorescence in response to \SI{222}{\nm} UV light from a KrCl excilamp has been reported in acrylic samples previously~\cite{acryl_cher}. The spectrum of the emission feature from~\cite{acryl_cher} does not match the spectrum indicated here. This may be an indication that the fluorescence is dependent on the excitation wavelength or the batch of acrylic. %(if the response in~\cite{acryl_cher} is from the Cherenkov light) 
%or to the type of incident radiation, if it is in response to interactions of the electrons themselves.

In future work, we will quantify the fluorescence of acrylic relative to TPB and study its time structure, by using a photomultiplier tube in place of a spectrometer.  We also plan to measure the fluorescence response of PTFE, as well as the response of both acrylic and PTFE at cryogenic temperatures and under excitation with photons down to \SI{130}{\nano\meter}.

%Since most of the Cherenkov light produced in acrylic is expected to be in the UV range\footnote{This is usually not a detectable wavelength range in experiments that use photomultipliers coupled to acrylic light guides, such as DEAP-3600.} \cite{acryl_cher}, a wavelength shift to the visible range, even with a low efficiency, could cause significant amount of light to reach the wavelength-sensitive region of the detector, producing an excess of detectable light and thus making simulations of the detector disagree with data. This could be a concern for experiments that use acrylic or other plastics (that could potentially fluoresce) surrounding their target, like DEAP-3600~\cite{DEAP}. \todo{should I also cite JUNO?}

%\appendix
%\section{Some title} Please always give a title also for appendices.

\acknowledgments

%\paragraph{Note added.} This is also a good position for notes added
We thank Marcin Ku{\'z}niak from AstroCeNT for providing the DEAP acrylic samples and Jochen Wieser from Excitech GmbH for the plasma cleaning of the samples.
This work has been funded in Canada by NSERC, CFI-LOF and ORF-SIF, as well as by the McDonald Institute.


\end{document}